\documentclass[11pt]{amsart}

\usepackage{hyperref}
\usepackage{amssymb,amsfonts,latexsym,amsmath,amsthm,wrapfig,graphicx}
\input xy
\xyoption{all}

\newtheorem{thm}{Theorem}[section]

\newtheorem{remark}[thm]{Remark}

\newtheorem{example}{Conclusion}

\def\la{\label}
\def\be{\begin{equation}}
\def\beq{\begin{equation}}
\def\eeq{\end{equation}}
\def\ee{\end{equation}}
\def\bea{\begin{eqnarray}}
\def\eea{\end{eqnarray}}
\def\p{\partial}

\begin{document}

\title[Surface response analysis]{Surface response analysis and determination of confidence regions for atmospheric CO$_2$: \\ a global warming study for U.S.A. data}

\author{Iuliana Teodorescu}
\author{Chris Tsokos}
\address{Statistics Department, University of South Florida, Tampa Florida}

\begin{abstract}
Starting from the atmospheric CO$_2$ measurements taken in Hawaii between 1959 and 2008, a quadratic model with interactions was fitted, using 5 attributable variables. Surface response analysis returned the eigenvalues and eigenvectors at the critical point, which turns out to be of mixed type, with two positive eigenvalues, one null, and the rest negative. From these data, it is derived that the confidence regions in two variables are of various types (elliptic, hyperbolic, and degenerate).  Based on these results we indicate how to determine two-dimensional confidence regions for statistically-significant variables which are relevant contributors to the atmospheric CO$_2$ emissions.
\end{abstract}
\maketitle


\section{Introduction} \label{sec:intro} 

This article presents a study of quadratic and interaction effects in the regression modeling of atmospheric CO$_2$ as a function of the known anthropogenic contributing factors. From the perspective of nonlinear modeling and optimal selection of attributable variables (from the set of all possible contributing factors), it is a continuation of the original study presented in \cite{model}. Once the optimal model is found, we carry out a novel type of analysis, through the surface response analysis and determination of the corresponding multi-dimensional confidence regions  in the parameter space.  

The proposed model that we are developing takes into consideration individual contributions and interactions along with higher order contributions if applicable. In developing the statistical model, the response variable is the CO$_2$ in the atmosphere and is given in unit parts per million (PPM). In the present analysis, we used real yearly data that has been collected from 1959 to 2008 for the continental United States. The air samples were collected at Mauna Loa Observatory, Hawaii. The CO$_2$ emission data was obtained from Carbon Dioxide Information Analysis Center (CDIAC). The analysis presented here consists of two parts: we first partially replicate the comprehensive study performed in \cite{model} in order to select the relevant variables and their interactions, and then we perform the surface-response analysis (nonlinear modeling) to the model obtained in \cite{model}. 

The data comes from Oak Ridge National Laboratory Division of U.S. Department of Energy. 
The air samples collected at Mauna Loa Observatory, Hawaii and the data unit is in ppmv.

\subsection{Goals and applications of the surface response analysis}

The goals of the surface response analysis for this problem are summarized below:

\begin{itemize}
\item we will identify the various ways in which second-order interactions between relevant attributable variables impact the emissions of CO$_2$ into the atmosphere, by performing a canonical decomposition of the quadratic part of the model. This will provide for us the relevant combinations of attributable variables (the canonical variables), and their respective effect (increasing, decreasing, or neutral) on the CO$_2$ emissions; 
\item depending on the different types of contributions at second-order level, we will classify and compute the various types of confidence regions, for pairs of canonical variables. The classification will produce confidence regions of elliptical, hyperbolic and degenerate types, whose specific geometric parameters we will compute;  
\item based on the specific quadratic model that we obtain, we will perform a quantitative analysis of the contributions for each canonical variable, by numerically comparing their effects, relevant for fluctuations of CO$_2$ emissions at the level of 2\% of the annual average (IPCC report and recommendations for 2020-2050, \cite{ipcc});
\item finally, we use the results of the analysis to make recommendations for optimal management of various attributable variables, both from the point of emission reductions, and from that of ``cap-and-trade" policies, in order to optimize the energy and industry requirements of a state (or country) with respect to carbon emissions restrictions;
\item we conclude with an outline of further studies and planned comparative analysis between United States and European Union CO$_2$ data.
\end{itemize}


\section{Regression analysis and model building}

\subsection{Prior studies and results}

In \cite{model}, the complete regression analysis for all the attributable variables and their interactions (including quadratic terms) was carried out. The study \cite{model} showed that only a subset of 5 variables are statistically relevant, and only five of their interaction terms (no self-interactions) contribute in a significant manner to the total variability of the data. The analysis performed included not only the model-building and regression component, but also data filtering, cross-validation and multiple measures of reliability and fitness. In the next section we reproduce some of this procedures, in order to arrive at a consistent and complete second-order model, the starting point of our surface response analysis. 

\subsection{Second-order model: parameter determination and validation}

One of the underlying assumptions to construct the model is that the response variable should follow Gaussian distribution. It is known \cite{model} that the CO$_2$ in the atmosphere does not follow the Gaussian distribution. 

Therefore, the Box-Cox transformation is applied to the CO$_2$ atmosphere data to filter the data to be normally distributed. After the Box-Cox filter, we retest the data and it shows our data will follow normal distribution; thus, we proceed to estimate the coefficients of the contributable variables for the transformed CO$_2$ atmosphere data.

We can proceed to estimate the approximate coefficients of the contributable variables for transformed CO$_2$ in the atmosphere and obtain the coefficient of all possible interactions. 

At the same time, we can determine the significant contributions of both attributable variables and interactions.
We begin with seven attributable variables as previously defined as $X_2, \ldots, X_8$ in the dataset (since the values listed as $X_1$ in the dataset are not relevant, $X_1$ being just the sum of all variables $X_i , i \ge 2$), and arrive by applying the stepwise forward selection procedure at a model with only five relevant variables (subsequently renamed $x_1, x_2, x_3, x_4, x_5$, corresponding to the original variables $X_2, X_3, X_5, X_6, X_8$), and fifteen 2$^{\mbox{\rm{\small{nd}}}}$ order interactions between each pair. We find that only five interactions are statistically relevant at $\alpha = 0.01$ level.

Thus the result of estimation becomes the quadratic model with interactions (fully consistent with the results of \cite{model}):

$$
[\widehat{CO}_2]^{-2.376} = \beta_0 + \sum_{i=1}^5 \beta_i x_i + \sum_{i \le j = 1}^5 \beta_{ij}x_i x_j, 
$$
where the measure for goodness-of-fit ($R^2 = 0.9973$ and the $p-$value less than 0.0001), as well as parameters $\{\beta_k \}$, are found from the  SAS output:
$x_1 = $ Gas Fuels, $x_2 = $ Liquid Fuels, $x_3 = $ Gas Flares, $x_4 = $ Cement, $x_5 = $ Bunker. Their corresponding coefficients determined by the stepwise 
 SAS procedure are: 
 
 \begin{center}
 \begin{table}[h!]
  \caption{Linear regression coefficients for attributable variables.}
 \begin{tabular}{| c || l | c | c | c | c | c |}
 \hline 
 ${\mbox{Variable}}$ & ${\mbox{Intercept}}$ & $x_1$ & $x_2$ & $x_3$ & $x_4$ & $x_5$ \\
 \hline 
\hline 
$   \Big [10^{17} \times\Big] \beta $  & $3.196\cdot 10^{8}$ & $-2.586$  & $-129.6$ &  $-1939$ &  $6922$ & $-896.1$  \\
 \hline 
 \end{tabular}
 \end{table}
\end{center}
The only non-zero interaction coefficients are obtained as follows (after rescaling by a global scale factor of $10^{-19}$): 
$$ \beta_{13} = -2.107, \,\, \beta_{23} = 5.593, \,\, \beta_{24} = - 2.559, \,\, \beta_{35} = -58.22, \,\, \beta_{45} = 20.49 $$

Therefore, we can write our model in matrix notation  (where prime denotes transposition) as 

\be \la{model}
Y = \beta_0 + \beta' \cdot X + X' \cdot B \cdot X,
\ee
with the obvious identifications 

$$X' = (x_1, \ldots, x_5),  \,\, \beta' = (\beta_1, \ldots, \beta_5), \,\, B_{ij} = B_{ji} = \frac{1}{2}\beta_{ij}\,\, (i < j).$$ 

More precisely, the vector of coefficients $\beta$ (up to a scale factor of $10^{17}$) and the symmetric matrix $B$ (up to a scale factor of $10^{19}$) have the forms:

$$
\beta = 
\left [ 
\begin{array}{c}
-2.586 \\ -129.6 \\  -1939 \\ 6922 \\ -896.1
\end{array}
\right ]
\, 
B = 
\left [ 
\begin{array}{ccccc}
0 & 0 & -1.0535 & 0 & 0 \\
0 & 0 & 2.7965 & -1.2795 & 0 \\
-1.0535 & 2.7965  & 0 & 0 & -29.11 \\
0 &  -1.2795 & 0  & 0 & 10.245 \\
0 & 0  & -29.11 & 10.245 & 0 \\
\end{array}
\right ]
$$

In order to perform the surface response analysis for this model, we must bring it to the simplest expression, by finding first its normal form and then its canonical decomposition.
Since these operations require inverting the matrix of second-order interactions, we need a preliminary calculation in order to determine  its eigenvalues and corresponding orthonormal eigenvectors. 

\section{Eingenvalue analysis of the second-order interactions matrix}

We recall that $\lambda_k, V_k$ ($k = 1, \ldots, 5$) are the eigenvalues and normalized eigenvectors of the  matrix $B$ if they solve the systems of linear equations:
$$
B \cdot V_k = \lambda_k V_k, \quad V'_k \cdot V_p = \delta_{kp}, 
$$
with $\delta_{ij}$ the Kronecker symbol, defined by $\delta_{ij} = 1$ if $i = j$ and $\delta_{ij} = 0$ otherwise. Then the matrix $B$ has the {\emph{principal-value decomposition}} (c.f. \cite[Appendix \S C]{west})
\be \la{pvd}
B = \sum_{k=1}^5 \lambda_k V_k V'_k.
\ee
Since the matrix $B$ has the form 
$$
B = 
\left [ 
\begin{array}{ccccc}
0 & 0 & a & 0 & 0 \\
0 & 0 & b & c & 0 \\
a & b  & 0 & 0 & d \\
0 &  c & 0  & 0 & e \\
0 & 0  & d & e & 0 \\
\end{array}
\right ], \quad a, b, c, d, e \in \mathbb{R},
$$
it follows from a general calculation that its eigenvalues are symmetric with respect to the origin:
$\lambda_{1,2} > 0, \lambda_3 = 0, \lambda_4 = -\lambda_2,  \lambda_5 = -\lambda_1$, so 
$$
\lambda_1 > \lambda_2 >   0 > \lambda_4  > \lambda_5. 
$$
More precisely, the eigenvalues of a matrix of this form are given by:
$$
\lambda_{1,5} = \pm \sqrt{\frac{s^2 + \sqrt{s^4 - 4 p^2}}{2}}, \quad 
\lambda_{2,4} = \pm \sqrt{\frac{s^2 - \sqrt{s^4 - 4 p^2}}{2}}, \quad 
\lambda_3 = 0,
$$
where $s^2 = a^2 + b^2 + c^2 + d^2 + e^2$ and $p^2 = a^2(c^2+e^2) + (be - cd)^2$.

Indeed, upon computing numerically the eigenvalues (using the SAS RSREG procedure \cite{SAS} or Mathematica's Eigensystem procedure), we arrive at 
 \be \la{eval}
\lambda_1 =  - \lambda_5 = 31.0277\times 10^{-19}, \,\, \lambda_2 = - \lambda_4 = 0.446626 \times 10^{-19},  \,\, \lambda_3 = 0,
\ee
up to the software numerical precision.  

Another general result is that the eigenvector corresponding to the null eigenvalue $\lambda_3 = 0$ has the the form
$$
V_3' =  \left (\frac{be - cd}{ac}x_5, -\frac{e}{c}x_5, 0, 0, x_5 \right ), \quad x_5 \in \mathbb{R},
$$
that is to say its third and fourth entries are identically zero. Specifically for our model, the normalized eigenvector $V_3$ becomes 
$$
V_3' =  (0.619629,-0.778849, 0, 0 ,-0.0972326).
$$

The other four orthogonal and normalized eigenvectors are found to be 
$$
V_1 = 
\left [ 
\begin{array}{c}
-0.022643 \\ 0.0697857 \\ 0.666881 \\-0.235096  \\ -0.70329
\end{array}
\right ],
\,\, 
V_2 = 
\left [ 
\begin{array}{c}
-0.554542 \\ -0.437981 \\ 0.235096 \\ 0.666881 \\ -0.0256057
\end{array}
\right ],
$$
$$
V_4 = 
\left [ 
\begin{array}{c}
0.554542 \\ 0.437981 \\ 0.235096 \\ 0.666881 \\ 0.0256057
\end{array}
\right ], 
\,\,
V_5 = 
\left [ 
\begin{array}{c}
0.022643 \\ -0.0697857 \\ 0.666881 \\ -0.235096 \\ 0.70329
\end{array}
\right ].
$$

Since $B\cdot V_3 = 0$, it is useful to decompose the vector $X$ into the component parallel to $V_3$, $X_{\parallel}$, and the component perpendicular to $V_3$, $X_{\perp}$:
\be \la{perp}
X = X_{\parallel} + X_{\perp}, \quad X_{\parallel} = (V_3'\cdot X) V_3, \quad X_{\perp}' \cdot X_{\parallel} = 0. 
\ee
Then, we also have 
\be \la{p2}
B\cdot X = B\cdot X_{\perp}, \quad X_{\parallel} = (0.619629x_1-0.778849 x_2-0.0972326x_5)V_3,
\ee
so we conclude that the ``neutral" component of $X$, $X_{\parallel}$ (associated with the zero eigenvalue $\lambda_3$), does not depend at all on the attributable variables $x_3$ and $x_4$, but only on the linear combination
\be \la{neutral}
z_3 := V_3' \cdot X =  0.619629x_1-0.778849 x_2-0.0972326x_5.
\ee 
We will return to this important fact when discussing applications in the last section.

\subsection{Canonical analysis of the quadratic model}

Let $B^-$ represent the symmetric generalized inverse of the matrix $B$ (\cite[Appendix \S C]{west}) 
$$
B^- = {\sum_{k}}^{'} \lambda_k^{-1} V_k V'_k, 
$$
where the ``primed" sum is taken only over non-zero eigenvalues (excluding $\lambda_3$ in our case). Then 
clearly from \eqref{p2}, 
$$
B^- \cdot V_3 = 0, \quad B^{-}\cdot X_{\parallel} = 0.
$$
Using the decomposition \eqref{perp}, the model \eqref{model} becomes 
$$
Y = \beta_0 +  (\beta'\cdot V_3) z_3 + \beta' \cdot X_{\perp} +  X'_{\perp}\cdot B \cdot X_{\perp}
$$
In order to  bring this expression to its normal form, we begin by shifting the variable $X$ by a constant term 
$$
\widehat{X} = X + \frac{1}{2}B^{-}\cdot \beta.
$$
\begin{remark} \la{rem1}
This transformation does not change the ``parallel" component since 
$$
V_3' \cdot \widehat{X} = V_3 ' \cdot X, \quad V_3' \cdot B^{-} = 0.
$$
\end{remark}
We obtain the model 
$$
{Y} = \beta_0 +  (\beta'\cdot V_3) z_3 + \beta' \cdot \widehat{X}_{\perp} - \frac{1}{4}\beta' \cdot B^{-}\cdot \beta + \widehat{X}_{\perp}'\cdot B \cdot \widehat{X}_{\perp}
- \beta' \cdot B\cdot B^{-} \widehat{X}_{\perp}, 
$$
where we have used the property $B^- \cdot B \cdot B^- = B^-$. Since $B\cdot B^{-} \widehat{X}_{\perp}  = \widehat{X}_{\perp}$, 
$$
{Y} = \beta_0  -\frac{1}{4}\beta' \cdot B^{-}\cdot \beta +  (\beta'\cdot V_3) z_3 + \widehat{X}'\cdot B \cdot \widehat{X},
$$
so  we are now working with the normal quadratic form $\widehat{X}'\cdot B \cdot \widehat{X}$.  
Using again \eqref{pvd}, the quadratic form $\widehat{X}'\cdot B \cdot \widehat{X}$ becomes
$$
\widehat{X}' \left ( \sum_{k=1}^5 \lambda_k V_k V'_k \right ) \widehat{X} = \sum_{k=1}^5 \lambda_k  (\widehat{X}'V_k)(V'_k \widehat{X}) = 
\sum_{k=1}^5 \lambda_k  |V'_k\cdot \widehat{X}|^2 = \sum_{k=1}^5 \lambda_k z_k^2,
$$
where we have introduced the {\emph{canonical coordinates}} 
\be \la{z}
z_k := V'_k \cdot \widehat{X}, \quad k = 1, 2, \ldots, 5.
\ee
We note that this coordinate change is consistent with \eqref{neutral} and Remark~\ref{rem1}.

To conclude, we have the canonical form of the model

\be \la{canonical}
Y = Y_0 +  (\beta'\cdot V_3) z_3 + \lambda_1 (z_1^2 - z_5^2) + \lambda_2 (z_2^2 - z_4^2),
\ee
and specifically for our data:
$$
Y -Y_0 =  186.47\times 10^{-17} z_3 + 31.03\times 10^{-19} (z_1^2 - z_5^2) + 0.45\times 10^{-19} (z_2^2 - z_4^2).
$$

In the following section we will determine the various types of confidence regions for pairs of variables for this model.
As a preliminary step in this procedure, we must first find the {\emph{stationary point}} of the model, defined generically 
as the point in attributable variables-space, where all the partial derivatives of the response
variable $Y$, with respect to each independent variable, are simultaneously equal to zero (also known as the critical
point or the zero-gradient point).  

For the quadratic model \eqref{model}, this condition becomes simply 
$$
\frac{\p Y}{\p x_k} = 0 \Rightarrow \beta' + 2X' \cdot B = 0 \Rightarrow B\cdot X = -\frac{1}{2} \beta
$$
Using \eqref{perp} and \eqref{p2}, the equation becomes 
$$
B\cdot X_{\perp} = -\frac{1}{2} \beta \Rightarrow X_{\perp} = -\frac{1}{2}B^{-}\cdot \beta \Rightarrow \widehat{X}_{\perp} = 0.
$$
Together with \eqref{z}, this gives for the stationary point $z_1 = z_2 =  z_4 = z_5 = 0$.

\begin{figure}
\begin{center}
 \includegraphics*[width=9.5cm]     {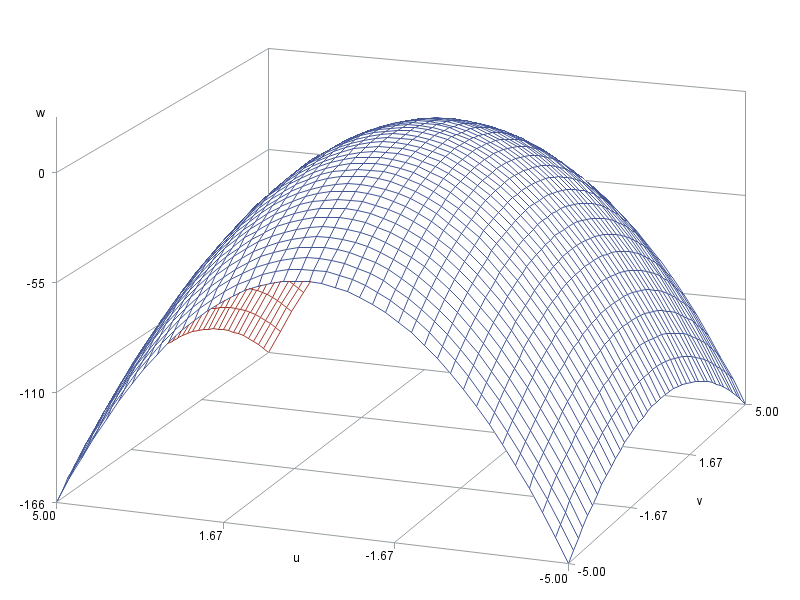}
 \includegraphics*[width=3cm]     {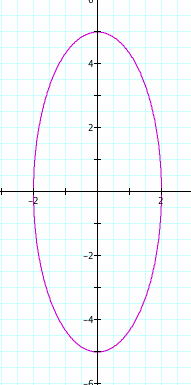}
\caption{Confidence regions for the elliptical case.}
\label{abs}
\end{center} 
\end{figure}

\subsection{Confidence region shapes and conic sections} 
In order to distinguish between various types of shapes the confidence regions may have, we now specialize to a pair of variables $(z_i, z_j) $ from the normal quadratic form written in canonical variables, and impose the inequality
$$
|Y - Y_0|  \le M, \quad M > 0,
$$
leading to
$$
\Big | \lambda_i z_i^2 + \lambda_j z_j^2 \Big | \le M, 
$$
which defines the confidence region centered at $(0, 0)$. We find the following cases, corresponding to classes of conic sections:

\subsubsection{Extremum point, elliptical region: all eigenvalues have the same sign}

If $\lambda_{i, j}$ are either all positive or all negative, the point $(0, 0)$ is a point of minimum or of maximum, respectively. The inequality becomes 
\be 
|\lambda_i| z_i^2 + |\lambda_j | z_j^2 \le M \Rightarrow \frac{z_i^2}{M/ |\lambda_i|} + \frac{z_j^2}{M/ |\lambda_j|} \le 1,
\ee
which defines the interior of an ellipse of semiaxes $\sqrt{M/ |\lambda_{i}|}, \sqrt{M / |\lambda_{j}|}$ (see Figure~\ref{abs}, right panel). 
The confidence region is given parametrically by:
\be \la{el}
z_i = \sqrt{\frac{M}{|\lambda_{i}|}} r \cos(\theta), \quad 
z_j = \sqrt{\frac{M}{|\lambda_{j}|}} r \sin(\theta), \quad 
0 \le r \le 1, \,\, 
\theta \in [0, 2\pi].
\ee

\begin{figure}[h!!!!!]
\begin{center}
 \includegraphics*[width=10cm]     {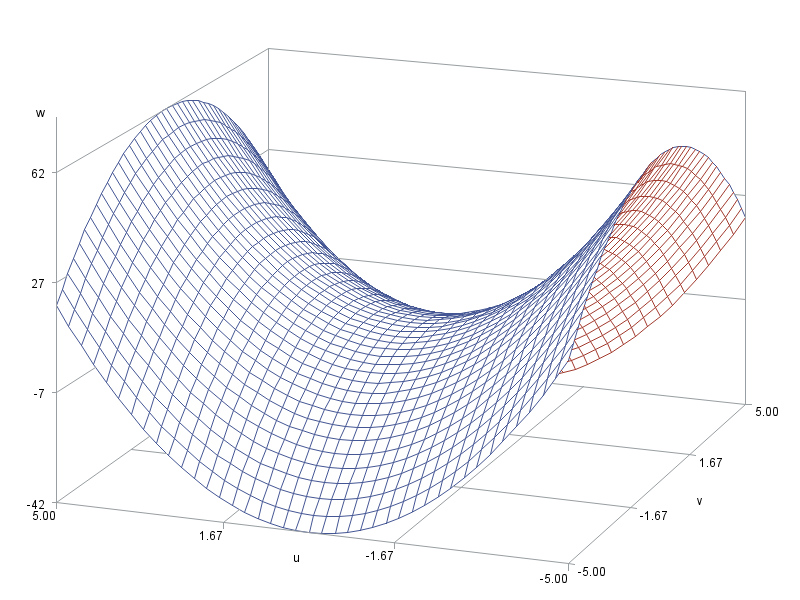}
 \includegraphics*[width=8cm]     {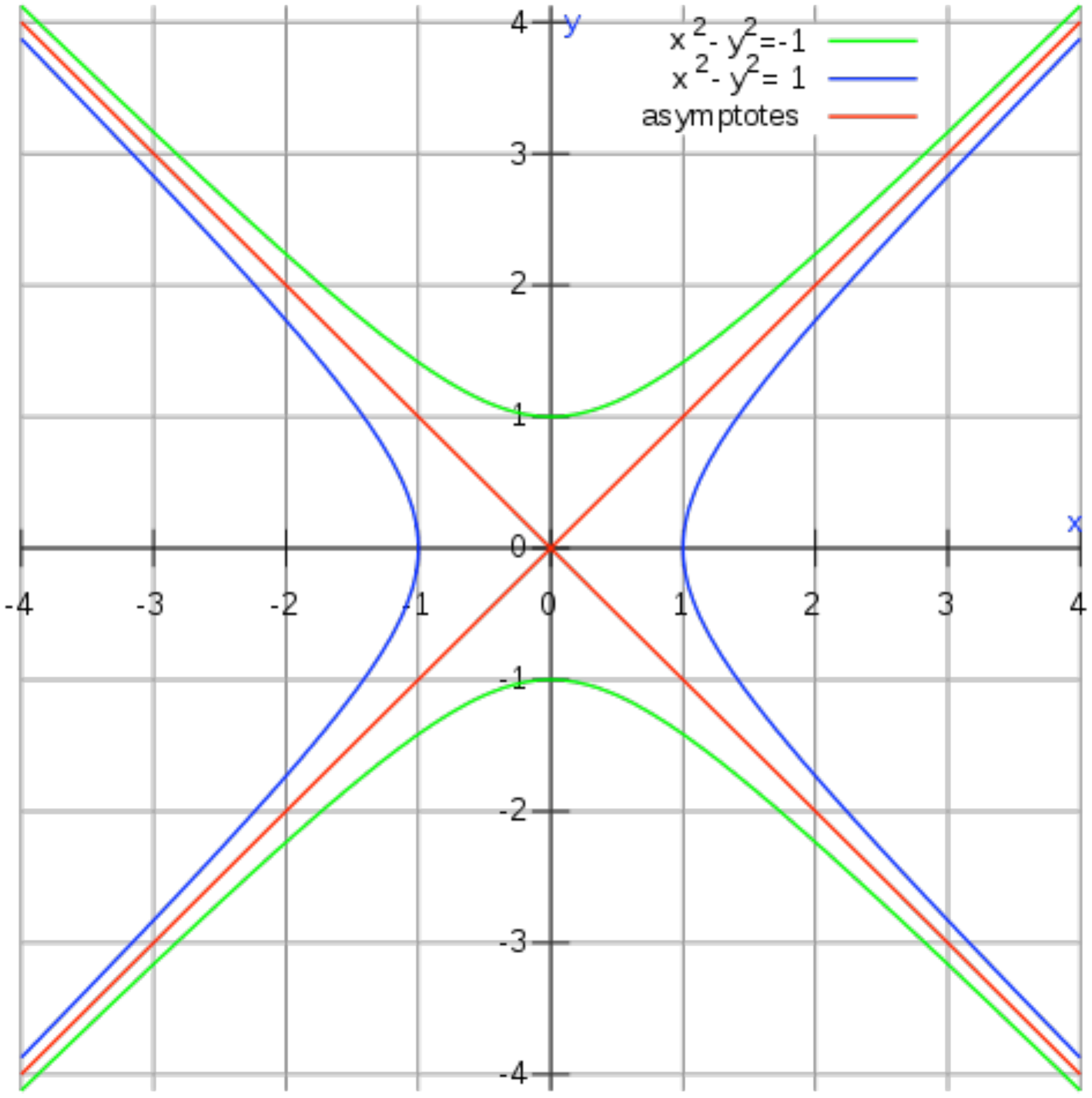}
\caption{Confidence regions for the hyperbolic case.}
\label{walker2}
\end{center} 
\end{figure}

\subsubsection
{Saddle-point, hyperbolic region: non-zero eigenvalues of different signs}

If, say, $\lambda_i > 0$ and $\lambda_j < 0$, then $(0, 0)$ is a saddle point, and  the inequality becomes 
$$
-M \le |\lambda_i| z_i^2 - |\lambda_j | z_j^2 \le M, 
$$
which defines the set of {\emph{orthogonal}} hyperbolas (see Figure~\ref{walker2})
\be 
\frac{z_i^2}{M/ |\lambda_i|} - \frac{z_j^2}{M/ |\lambda_j|} \le 1, \quad 
\frac{z_j^2}{M/ |\lambda_j|} - \frac{z_i^2}{M/ |\lambda_i|} \le 1.
\ee
The intersection of these conditions defines a region that looks like an elongated rectangle (elongated ``corners", the domain defined by the blue and green curves in Figure~\ref{walker2}) and can be approximated with a rectangular shape.
The confidence region is given parametrically by:

\be \la{hy}
z_i = \sqrt{\frac{M}{|\lambda_{i}|}} r \cosh(t), \quad 
z_j = \sqrt{\frac{M}{|\lambda_{j}|}} r \sinh(t), \quad 
-1 \le r \le 1, \,\, 
t \in \mathbb{R}.
\ee

\subsubsection{Flatness point, conical region: some eigenvalues are zero (degenerate)}

\begin{figure}[h!!!!!]
\begin{center}
 \includegraphics*[width=10cm]     {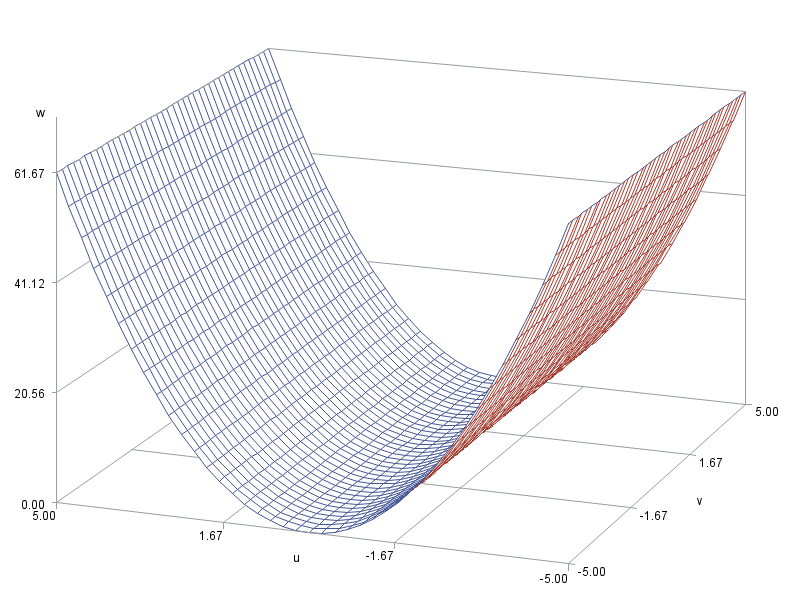}
\caption{Confidence regions for the  degenerate (flatness) case.}
\label{deg}
\end{center} 
\end{figure}

Let now $\lambda_j \to 0$ in the previous case, and we obtain a poinf of ``flatness" or degenerate point, where the inequality becomes
$$
\lambda_i z_i^2 \le M, \quad z_j \in \mathbb{R},
$$
which corresponds to the conical degeneration of a hyperbolic region (the domain defined by the blue and red curves in Figure~\ref{walker2}), i.e. an infinite strip domain, Figure~\ref{deg}: 
\be \la{co}
|z_i| \le \sqrt{\frac{M}{|\lambda_i|}}
\ee

%
%
%

\section{Applications: predictions based on nonlinear analysis} 

\subsection{Order of magnitude analysis for the canonical variables}

We begin addressing the list of applications described in the Introduction by first providing numerical estimates  for each term in the specific model derived earlier:
$$
Y -Y_0 =  186.47\times 10^{-17} z_3 + 31.03\times 10^{-19} (z_1^2 - z_5^2) + 0.45\times 10^{-19} (z_2^2 - z_4^2).
$$
From the data, as well as from the estimate of the response variable at the stationary point, we obtain (at order of magnitude) the following estimate for $Y_0$:
$$
Y_0 \sim ({\mbox{CO}}_2)^{-2.376} \sim O( 10^{-7}) 
$$
Using the IPCC recommendation for CO$_2$ emissions reduction \cite{ipcc}, of 20\% -- 30\% through 2020, we obtain an annual variation of the order of 3\%, which means a variation of the order of 10\% for the response variable $Y$ (note  that increasing  CO$_2$ corresponds to  {\emph{decreasing}}  $Y$). Therefore, it is reasonable and relevant to work with variations of the order $M \sim 0.1\times Y_0  = O(10^{-8})$. 

At this order of magnitude, a simple estimate for the variation of the canonical variable $z_3$ gives us the value of $|z_3| \sim 10^{15} \times M = O(10^7)$. By contrast, applying the formulas \eqref{el}, \eqref{hy}, \eqref{co} and the numerical values for $\lambda_{1, 2}$, we obtain the order of magnitude 
$$
|z_k| \sim O\left ( \sqrt{\frac{M}{|\lambda_k|} } \right ) \sim O(10^5), \quad k = 1, 2, 4, 5.
$$
This indicates that, while the canonical variable $z_3$ \eqref{neutral} may be allowed to fluctuate up to order $10^7$ around the origin, the other canonical variables are much more restricted, by up to 2 orders of magnitude less. Since the variable $z_3$ does not contain any contribution from the attributable variables $x_3, x_4$, this analysis singles them out in a two-fold way: 
\begin{itemize}
\item[(i)] their variation (no matter how small) always contributes to the quadratic part of the response variable, and 
\item[(ii)] the order of magnitude allowed for their variations, at given threshold $M \sim 0.1 \times Y_0$, is about 100 times smaller than what is allowed for the linear combination 
$z_3$ \eqref{neutral}.
\end{itemize}

Therefore, we arrive at the following conclusion with direct practical applications: 
\begin{example}
For variations of the CO$_2$ levels at the order of magnitude stipulated by IPCC (around 2\% per year), the linear combination of 
attributable variables $z_3 =  0.619629x_1 - 0.778849x_2 - 0.0972326x_5$ can be considered to be basically ``free" compared to the 
other canonical variables, i.e. it may have fluctuations of up to order $O(10^6)$ {\bf{without having a significant effect}} on the CO$_2$ levels.
\end{example}

Moreover, we can estimate the order of magnitude of $M$ at which the variable $z_3$ stops being ``free" with respect to the other variables, from the simple comparison
$$
10^{15} \times M \sim O\left ( \sqrt{\frac{M}{|\lambda_k|} } \right ) = 10^9 \times \sqrt{M} \Rightarrow M \sim 10^{-12} \Rightarrow \frac{M}{Y_0} \sim 10^{-5} = 0.001\%
$$
In other words, unless we are concerned with yearly variations of the CO$_2$ levels not exceeding $0.001\%$ of the current levels (an accuracy not realistic for our present 
measurement and prediction capabilities), Conclusion 1 holds.

\subsection{Managing CO$_2$ emissions: accountability  policies and metrics}

Throughout this subsection, we let the values of the attributable variables $X' = (x_1, x_2, x_3, x_4, x_5)$ be measured from the stationary point 
$X_s = -\frac{1}{2}B^{-}\cdot \beta$. In other words, instead of $x_1$ we use the shifted value $x_1 + \left (\frac{1}{2}B^{-}\cdot \beta  \right )_1$, instead of $x_2$ we use the shifted value $x_2 + \left (\frac{1}{2}B^{-}\cdot \beta  \right )_2$, etc.

Starting from the model 
$$
Y -Y_0 =  186.47\times 10^{-17} z_3 + 31.03\times 10^{-19} (z_1^2 - z_5^2) + 0.45\times 10^{-19} (z_2^2 - z_4^2),
$$
and the defining relations for the linear combinations
$$
\begin{array}{lll}
z_1 & =  & -0.022643x_1 + 0.0697857x_2 + 0.666881x_3 -0.235096x_4 - 0.70329x_5, \\
z_2 & = & -0.554542x_1 -0.437981x_2 + 0.235096x_3 + 0.666881x_4  -0.0256057x_5, \\
z_4 & = &  0.554542x_1 + 0.437981x_2 + 0.235096x_3 + 0.666881x_4 + 0.0256057x_5, \\
z_5 & = & 0.022643x_1 - 0.0697857x_2 + 0.666881x_3 -0.235096x_4 + 0.70329x_5, 
\end{array}
$$
and using Conclusion 1 (which allows to neglect the term proportional to $z_3$ from the model), we arrive at the following equation:

$$
M \simeq 124.12\times 10^{-19} u_1 v_1 + 1.8\times 10^{-19} u_2 v_2, 
$$
where 
$$
u_1 = 0.666881x_3 -0.235096x_4, \quad u_2 = 0.235096x_3 + 0.666881x_4, 
$$
$$
v_1 =  -0.022643x_1 + 0.0697857x_2 - 0.70329x_5, 
$$
$$
v_2 = -0.554542x_1 -0.437981x_2 -0.0256057x_5,
$$
which together with $z_3$ form a new set of orthogonal coordinates in $\mathbb{R}^5$ (just like $\{x_k\}$ and $\{ z_k \}$). In order to implement 
a constraint at given value of $M$, we may choose to set either the product $u_1 v_1 = 0$ or $u_2 v_2 = 0$, and solve for the remaining term. This 
choice will provide a direct procedure for comparing the relative weight of one attributable variable versus another. 

\subsubsection{Example} We choose to set $u_1 = 0$, which leads to the conclusion that a variation of 1 unit in the attributable variable $x_3$ 
is offset by a variation of $0.666881/0.235096 \simeq 2.837$ units in the variable $x_4$. The new variable $u_2$ now becomes 
$$
u_2 = (0.235096 + 0.666881\cdot 2.837) x_3 \simeq 2.127 x_3
$$
 Choosing $M \sim 10^{-8}$ again, we obtain the inequality
$$
|u_2 v_2| \le 5.5 \times 10^{10} \Rightarrow |v_2| \le \frac{2.586}{|x_3|} 10^{10}. 
$$
Specifically, consider the situation where we wish to increase the value of $x_3$ (Gas Flares) by $10^3$. From the previous analysis, in order for the
total emissions not to exceed 2\% of yearly average ($M \sim 10^{-8}$), an increase of 1000 units in Gas Flares may be accompanied by an increase in $x_4$ (Cement) 
of 2127 units, while the linear combination $v_2$ must satisfy
$$
|0.554542x_1  + 0.437981x_2  + 0.0256057x_5| \le 25.86 \times 10^6.
$$
The values of the linear combinations $z_3, v_1$ remain {\emph{arbitrary}} in this case:
$$
v_1 =  -0.022643x_1 + 0.0697857x_2 - 0.70329x_5 \in \mathbb{R},
 $$
 $$ z_3 =  0.619629x_1 - 0.778849x_2 - 0.0972326x_5 \in \mathbb{R}.
$$
\begin{example}
By performing the canonical decomposition for a model with interactions as described here, we can identify linear combinations 
of attributable variables (such as $u_1$) which lead to a rigorous ``trade value" of one variable versus another, in the context of 
a given ``cap" on the level of total emissions ($M$). The analysis provides an explicit procedure for the {\bf{strategy of trading of attributable 
variables}}, which may be use to accommodate industrial requirements, while obeying CO$_2$ emission limitations. 
\end{example}

\section{Further studies and general remarks}

\begin{itemize}
\item In forthcoming publications, we will present a similar analysis to the present article and the initial work \cite{model}, where we will determine the optimal second-order model with interactions and subsequent surface response analysis, in the case of European Union atmospheric CO$_2$ data. A comparative analysis will then be carried out, with the purpose of identifying similarities and discrepancies between the US and EU models;
\item In the process of devising an effective way of approximating individual confidence intervals for the original attributable variables, we discovered a novel approach based on recent results stemming from real-algebraic geometry and polynomial optimization for matrix-valued variables. The new method allows to distinguish extreme scenarios (best case and worst case) by making use of a ``matrix sum-of-squares" decomposition in a convex cone of $\mathbb{R}^n$. Given the complexity of the mathematical framework involved, we will present the new method, applied to problems of the type discussed here, in a separate article \cite{alg}.
\end{itemize}

\section{Acknowledgments}

The authors wish to thank R. Teodorescu for pointing out the relation between estimating high-dimensional confidence regions and the real-algebraic methods to be detailed in \cite{alg}.

\end{document}